\documentclass{jpsj3}  
\usepackage{txfonts}

\title{Flux Phase in Bilayer $t-J$ Model: 
Time-Reversal Symmetry Breaking Surface State without 
Spontaneous Magnetic Field
}

\author{Kazuhiro Kuboki \thanks{kuboki@kobe-u.ac.jp}}
\inst{Department of Physics, Kobe University, Kobe 657-8501, Japan
} 

\abst{
We study surface states of high-$T_C$ cuprate superconductor 
YBCO using the bilayer $t-J$ model. Calculations based on the 
Bogoliubov de Gennes method show that a flux phase that breaks 
time-reversal symmetry (${\cal T}$) may arise near a (110) 
surface where the $d_{x^2-y^2}$-wave superconductivity is 
strongly suppressed. It is found that the flux phase in which spontaneous 
magnetic fields in two layers have opposite directions may be stabilized 
in a wide region of doping rate, and split peaks in the local density 
of states appear.  Near the surface, spontaneous magnetic field
may not be observed experimentally, because the contributions 
from two layers essentially cancel out. This may explain the absence 
of local magnetic field near the (110) surface of YBCO, for which 
the sign of ${\cal T}$ violation has been detected.

 }

\begin{document}
\maketitle

\section{Introduction}

In high-$T_C$ cuprate superconductors, spontaneous violation of 
time-reversal symmetry (${\cal T}$) has been observed in various 
kinds of experiment.\cite{Coving,TRSB1,TRSB2,TRSB3} 
One of the famous example is the peak splitting of zero bias conductance 
in ab-oriented YBCO/insulator/Cu junction.\cite{Coving}  
This has been interpreted as a consequence of the occurrence of 
second superconducting (SC) order parameter (OP) near the surface, 
which has symmetry different from 
that in the bulk.\cite{Fogel,Matsu1,Matsu2} 
For this type of surface state, 
spontaneous current would flow along the surface,  
and a magnetic field should be generated locally.
However, experimental evidence for such magnetic fields 
is still controversial.\cite{Carmi,msr}

The present author has studied the (110) surface state of high-$T_C$ 
cuprates based on the Bogoliubov-de Gennes (BdG) method applied to a 
single-layer $t-J$ model, and found that a different kind of 
${\cal T}$-breaking surface state, flux phase,  can occur.\cite{KK1,KK2}
The flux phase is a mean-field solution to the $t-J$ model in which 
staggered currents flow and the flux penetrates a plaquette in a square 
lattice.\cite{Affleck} This state has free energy higher than that of 
the $d_{x^2-y^2}$-wave SC state except very near half filling, 
so that it is only a metastable state in uniform 
systems.\cite{Zhang,Hamada,Bejas,Zhao}  
Near (110) surfaces $d_{x^2-y^2}$-wave SC state is strongly 
suppressed, and then the flux phase may arise locally leading to a 
${\cal T}$-breaking surface state. 
However, the doping region in which ${\cal T}$ violation occurs 
was much narrower than that observed experimentally in YBCO,  
if we use an effective single-layer model.\cite{KK1,KK2} 

Later reexamination using a bilayer $t-J$ model that describes the electronic 
states of YBCO more accurately have shown that the flux phase may 
occur as a metastable state in a doping region much wider than 
that for the effective single-layer model.\cite{KK3} 
For the bilayer $t-J$ model, there may be two types of flux phase 
in which the directions of the flux in two layers are the same or opposite, and 
a phase transition occurs from the latter to former
as the doping rate 
increases.\cite{KK3}
We call the former (latter) one as a type A (B) flux phase. 
If the type B flux phase occurs near the (110) surface, 
the spontaneous magnetic field should be very small, 
since the contributions from two layers essentially cancel out.
This may explain why no magnetic field is observed in some experiments 
for the (110) surface state of YBCO.

In this paper, we study the (110) surface states of YBCO system 
that are described by the bilayer $t-J$ model. 
Spatial variations of the OPs are treated using the 
BdG method\cite{dG}, and we will show that the flux phase can occur 
in a wide region of the doping rate when the SC order is suppressed. 
The local density of states (LDOS) is also examined to see whether 
the splitting of the zero-energy peak occurs in agreement 
with experimental results. 

This paper is organized as follows. In Sect. 2 the model is presented 
and the BdG equations are derived. Results of numerical calculations 
are described in Sect. 3, and Sect. 4 is devoted to summary.

\section{Bogoliubov de Gennes Equations}

We consider the bilayer $t-J$ model on a square lattice 
whose Hamiltonian is given by $H  =  H_1 + H_2 + H_\perp$ with  
\begin{eqnarray}
\displaystyle 
H_i =&\displaystyle -\sum_{j,\ell,\sigma} 
t_{j\ell} {\tilde c}^{(i)\dagger}_{j\sigma} {\tilde c}^{(i)}_{\ell\sigma}
 +J\sum_{\langle j,\ell\rangle} {\bf S}^{(i)}_j\cdot {\bf S}^{(i)}_\ell, 
 \ \ \  (i=1,2)\\
 H_\perp =&\displaystyle -\sum_{j,\ell,\sigma} t_{j\ell}^\perp 
\Big( {\tilde c}^{(1)\dagger}_{j\sigma} {\tilde c}^{(2)}_{l\sigma} + h.c.\Big)
 +J_\perp\sum_j {\bf S}^{(i1}_j\cdot {\bf S}^{(2)}_j, 
\end{eqnarray}
where the transfer integrals (in plane) $t_{j\ell}$ are finite for the first-  ($t$), 
second-  ($t'$), and third-nearest-neighbor bonds ($t''$), or zero otherwise.  
$J$ $(J_\perp)$ is the inplane (interplane) antiferromagnetic superexchange
 interaction, and $\langle j,\ell \rangle$ denotes nearest-neighbor bonds.\cite{Ogata} 
The interplane transfer integrals $t^\perp_{j\ell}$ are chosen to reproduce 
the dispersion in $k$ space,\cite{Andersen} 
$t^\perp_k =  -t^\perp_0 (\cos k_x - \cos k_y)^2$, namely, 
"on-site" ($t^\perp_0$), second- ($t^\perp_2 = -t^\perp_0/2$) , 
and third-nearest-nearest-neighbor bonds ($t^\perp_3 = t^\perp_0/4$) 
are taken into account. 

${\tilde c}_{j\sigma}^{(i)}$ is the electron operator for the $i$-th plane 
in Fock space without double occupancy, and we treat this condition 
using the slave-boson method\cite{Ogata,Zou,Lee}   
by writing ${\tilde c}_{j\sigma}^{(i)}=b_j^{(i)\dagger} f_{j\sigma}^{(i)}$ 
under the local constraint 
$\sum_{\sigma}f_{j\,\sigma}^{(i)\dagger}f_{j\,\sigma}^{(i)} 
+ b_j^{(i)\dagger}b_j ^{(i)}= 1$ 
at every $j$ site. Here $f_{j\sigma}^{(i)}$ ($b_j^{(i)}$) is a fermion (boson) operator  
that carries spin $\sigma$ (charge $e$); the fermions (bosons) are frequently 
referred to as spinons (holons). 
The spin operator is expressed as 
$
 {\bf S}_j ^{(i)} = \frac{1}{2}\sum_{\alpha,\beta}
f^{(i)\dagger}_{j\alpha} {\bf \sigma}_{\alpha\beta}f_{j\beta}^{(i)}$. 

We decouple the Hamiltonian in the following 
manner.\cite{Kotliar,Suzumura} 
The bond OPs  in plane 
$ \sum_\sigma \langle f^{(i)\dagger}_{j\sigma}f_{l\sigma}^{(i)} \rangle$ 
and $ \langle b^{(i)\dagger}_j b_l^{(i)} \rangle$ 
are introduced, and we denote 
$\chi_{jl}^{(i)} \equiv  
\sum_\sigma \langle f^{(i)\dagger}_{j\sigma}f_{l\sigma}^{(i)} \rangle$
for nearest-neighbor bonds. 
The interlayer bond OP is defined as 
$\chi^\perp_j \equiv \sum_\sigma \langle f^{(1)\dagger}_{j\sigma}
f^{(2)}_{j\sigma} \rangle$.  
Although the bosons are not condensed in purely two-dimensional systems  
at finite temperature ($T$), they are almost condensed at a low $T$ and for 
finite carrier doping ($\delta \gtrsim 0.05$). 
Since we are interested in the low temperature region
 ($T \leq 10^{-2}J \sim 10$K),   
we treat holons as Bose condensed.
Hence, we approximate $ \langle b_j^{(i)}\rangle \sim \sqrt{\delta}$ and 
$ \langle b^{(i)\dagger}_j b_l^{(i)} \rangle 
\sim \delta$ ($\delta$ being the doping rate), 
and replace the local constraint with a global one, 
$\frac{1}{N}\sum_{j,\sigma} \langle f^{(i)\dagger}_{j\sigma} f_{j\sigma}^{(i)}\rangle 
= 1-\delta$,  where $N$ is the total number of lattice sites within a plane. 
This procedure amounts to renormalizing  the transfer integrals 
by multiplying $\delta$, 
{\it e.g.}, $t \to t\delta$, {\it etc.}, 
and rewriting ${\tilde c}_{j\sigma}^{(i)}$ as $f_{j\sigma}^{(i)}$. 
In a qualitative sense, this approach is equivalent to 
the renormalized mean-field (MF) theory of Zhang {\it et al.}\cite{Zhang2} 
(Gutzwiller approximation). 
The spin-singlet resonating-valence-bond (RVB) OP  on the bond $\langle j,l\rangle$ 
 is given as 
$\Delta_{j,l}^{(i)} = \langle f_{j\uparrow}^{(i)}f_{l\downarrow}^{(i)} 
-f_{j\downarrow}^{(i)}f_{l\uparrow}^{(i)}\rangle/2$. 
The interlayer RVB OP is defined as 
$\Delta^\perp_j \equiv \langle f^{(1)}_{j\uparrow}f^{(2)}_{j\downarrow}
- f^{(1)}_{j\downarrow}f^{(2)}_{j\uparrow}\rangle/2$. 
 Under the assumption of the Bose condensation of holons, $\Delta_{j,l}$ 
 is equivalent to the SCOP.

We treat a system with a (110) surface, and 
denote the direction perpendicular (parallel) to the (110) surface as $x$ ($y$). 
The $x$ coordinate is given as $x_j= j_xa$  where 
$a=a'/\sqrt{2}$ with $a'$ being the lattice constant of the square lattice.
In order to describe the Flux phase and the SC state, 
$\chi^{(i\pm)}_j \equiv \chi^{(i)}_{j,j+x\pm y}$ and 
$\Delta^{(i\pm)}_j \equiv \Delta^{(i)}_{j.j+x\pm y}$ are defined for the 
$i$-th plane. 
We assume that the system is uniform along the $y$ direction, 
and consider the spatial variations of OPs only in the $x$ direction. 
By imposing the periodic boundary condition for the $y$ direction, 
the Fourier transformation for the $y$ coordinate is 
performed.\cite{KKBdG,Tanuma,Zhu,KKBdG2}
(Hereafter we write $j_x$ simply as $j$, and take $a=1$.)
Then the MF Hamiltonian is written as follows 
\begin{equation}
\displaystyle 
{\cal H}_{ MF} = \sum_k\sum_{j,l}\Psi_j^\dagger(k) 
{\hat h}_{jl}(k) \Psi_l(k), 
\end{equation} 
with $\displaystyle \Psi_j^\dagger(k) 
= \big(f_{j\uparrow}^{(1)\dagger}(k), f_{j\downarrow}^{(1)}(-k), 
f_{j\uparrow}^{(2)\dagger}(k), f_{j\downarrow}^{(2)}(-k)\big)$, and 
$k$ is the wave number along the $y$ direction. 
The matrix ${\hat h}_{ij}(k)$ is given as
\begin{equation}
\displaystyle {\hat h}_{jl}(k) = 
\left (\begin{array}{cccc}
\xi^{(1)}_{jl}(k)  & F^{(1)}_{jl}(k) & \epsilon_{jl}(k) & f_{jl} \\
F_{lj}^{{(1)}*}(k) & -\xi^{(1)}_{lj}(-k) & f^*_{jl} & -\epsilon_{lj}^*(k) \\
\epsilon_{lj}^*(k) & f_{jl} & \xi^{(2)}_{jl}(k) &  F^{(2)}_{jl}(k) \\
f_{jl}^* & -\epsilon_{jl}(k) & F_{lj}^{{(2)}*}(k) &  -\xi^{(2)}_{lj}(-k)
\end{array}\right ), 
\end{equation}
where
\begin{equation}\begin{array}{rl}
\xi^{(i)}_{jl}(k) = & \displaystyle 
-\delta_{j,l}(\mu +2t'\delta \cos 2k) \\ 
& \displaystyle -\delta_{j,l-1} 
\big[2t\delta \cos k  + \frac{3J}{8}
( \chi^{(i+)}_{j}e^{ik} + \chi^{(i-)}_{j}e^{-ik}) \big] \\
& \displaystyle  -\delta_{j,l+1}
\big[2t\delta \cos k + \frac{3J}{8}
 ((\chi^{(i+)}_{l})^*e^{-ik} + (\chi^{(i-)}_{l})^*e^{ik}) \big] \\
& \displaystyle -(\delta_{j,l-2}+\delta_{j,l+2})(t' + 2t''\cos 2k)\delta,  
\\ 
& \\
\epsilon_{jl}(k) = & \displaystyle 
-\delta_{jl}\big[(t^\perp_0 + 2t^\perp_2 \cos 2k)\delta 
+ \frac{3J_\perp}{8}(\chi^\perp_j)^*\big]
\\
& \displaystyle -(\delta_{j,l-2}+\delta_{j,l+2})
(t^\perp_2 + 2t^\perp_3 \cos 2k)\delta,
\\
& \\
F^{(i)}_{jl}(k) = & \displaystyle  \frac{3J}{4} 
\big[\delta_{j,l-1}(\Delta^{(i+)}_j e^{ik} + \Delta^{(i-)}_j e^{-ik}) \\
& \displaystyle + \delta_{j,l+1} (\Delta^{(i+)}_l e^{-ik} + \Delta^{(i-)}_l e^{ik})\big],    
\\
& \\
f_{jl} = & \displaystyle \delta_{jl}\frac{3J_\perp}{4}\Delta^\perp_j, 
\end{array}\end{equation}
with $\mu$ being the chemical potential, 

We diagonalize the MF Hamiltonian by solving the
following BdG equation for each $k$, 
\begin{equation} 
\sum_l {\hat h}_{jl}(k) u_{ln}(k) 
= E_n(k) u_{jn}(k), 
\end{equation} 
where $E_n(k)$ and $u_{jn}(k)$ are the energy eigenvalue
and the corresponding eigenfunction, respectively, for each
$k$. The unitary transformation using $u_{jn}(k)$ diagonalizes
the Hamiltonian ${\cal H}_{MF}$, and the OPs and the spinon number 
at the $j$ site for the layer 1 can be obtained as, 
\begin{equation}\begin{array}{rl}
\langle n^{(1)}_j\rangle =  & \displaystyle 
\frac{1}{N_y} \sum_{k,n} 
\Big[\big| u_{4j-3,n}(k) \big|^2 f(E_n(k))
\\ 
 & \displaystyle + \big| u_{4j-2,n}(k) \big|^2 \big[1-f(E_n(k))\Big],  
\\
\chi^{(1\pm)}_j = 
& \displaystyle \frac{1}{N_y} \sum_{k,n} 
\Big[u_{4j+1,n}^*(k)u_{4j-3,n}(k)e^{\mp ik}f(E_n(k))
\\
&  \displaystyle + u_{4j+2,n}(k) u_{4j-2,n}^*(k)e^{\pm ik}(1-f(E_n(k)))\Big], 
\\
\Delta^{(1\pm)}_j = & \displaystyle \frac{1}{4N_y} \sum_{k,n} 
\Big[u_{4j-3,n}(k)u_{4j+2,n}^*(k)e^{\mp ik} 
\\
& \displaystyle + u_{4j-2,n}^*(k)u_{4j+1,n}(k)e^{\pm ik}\big]
\tanh\Big(\frac{E_n(k)}{2T}\Big)\Big], 
\end{array}\end{equation}
where $N_x$ ($N_y$) and $f$ are the number of lattice sites along 
the $x$ ($y$) direction within a plane, and the Fermi distribution function, 
respectively. The OPs and the spinon number for the layer 2 are obtained by replacing 
the subscripts, $(4j-3) \to (4j-1)$, $(4j-2) \to (4j)$, {\it etc.},  
in Eq. (7). 
The interlayer OPs are given as, 
\begin{equation}\begin{array}{rl}
\chi^\perp_j = & \displaystyle \frac{1}{N_y} \sum_{k,n}
\Big[u_{4j-3,n}^*(k)u_{4j-1,n}(k)f(E_n(k)) 
\\
& \displaystyle + u_{4j-2,n}(k) u_{4j.n}^*(k)(1-f(E_n(k)))\Big], 
\\
\Delta^\perp_j = & \displaystyle \frac{1}{4N_y} \sum_{k,n} 
\Big[u_{4j-3,n}(k)u_{4j,n}^*(k) + u_{4j-2,n}^*(k) u_{4j-1.n}(k)\Big]
\\ 
& \displaystyle \times \tanh\Big(\frac{E_n(k)}{2T}\Big).
\end{array}\end{equation}

The $d$- and $s$-wave SCOPs are obtained by combining 
$\Delta^{(i\pm)}$s: 
$\Delta^{(i)}_d(j) = (\Delta^{(i+)}_j - \Delta^{(i-)}_j + \Delta^{(i+)}_{j-1} 
- \Delta^{(i-)}_{j-1})/4$ and 
$\Delta^{(i)}_s(j) = (\Delta^{(i+)}_j + \Delta^{(i-)}_j + \Delta^{(i+)}_{j-1} 
+ \Delta^{(i-)}_{j-1})/4$.

\section{Surface States and Local Density of States}
In this section we present the results of numerical calculations for surface states. 
The procedure of numerical calculations 
is the following. We diagonalize the Hamitotonian 
${\cal H}_{MF}$ with the OPs substituted in matrix elements, 
and the resulting eigenvalues and eigenfunctions 
are used to recalculate the OPs. This procedure is iterated until the 
convergence is reached. 
For the system size, $N_x=200$ and $N_y=100$ are used throughout. 
The band parameters are chosen after Ref. 30;  
$t/J=2.5$, $t'/t=-0.3$, $t''/t=0.15$, $t^\perp_0/t=0.15$, and $J_\perp/J=0.1$.  
These parameters were chosen to reproduce experimental results for 
YBCO.\cite{Yamase} 
We restrict ourselves to the case of 
low temperature, $T= 10^{-3}J$ ($\sim 1$K). 

The spatial variations of the OPs for $\delta=0.15$ are shown in Figs. 1 and 2. 
It is seen that the $d$-wave SCOP is suppressed near a (110) surface. 
The imaginary parts of the bond OPs (Im $\chi$) are finite there, 
and Im $\chi$ for different layers have opposite signs.
This means that the flux phase arises leading to a ${\cal T}$-breaking surface state. 
Spontaneous current flowing  along the surface is given as, 
\begin{equation}
\displaystyle 
J_y^{(i)}(j) = \frac{\sqrt{2} \pi t \delta}{\phi_0} 
{\rm Im} \ \chi^{(i+)}_j, 
\end{equation} 
with $\phi_0=h/2e$ being the flux quantum. 
(In principle, there is a term proportional to the vector potential in $J_y$, 
but we neglect it for simplicity.) 
From this equation, we see that the directions of the currents and those of 
the flux in two layers are opposite (type B flux phase). 
In this case, the spontaneous magnetic field 
near the surface will be very small, since the contributions from 
two layers essentially cancel out.
Then it will be hard to observe it experimentally. 
Small imaginary parts of the $s$-wave SCOP (Im $\Delta_s$) also appear 
near the surface, and their signs in two layers are also opposite. 
%
%
%
\begin{figure}[htb]
\begin{center}
\includegraphics[width=6.0cm,clip]{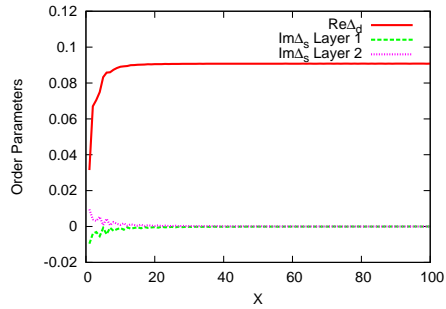}
\caption{(Color online) Spatial variations of the SCOPs for $\delta=0.15$. 
Here $x$ is measured in units of lattice spacing $a$. 
Note that all OPs are nondimensional.}
\end{center}
\end{figure}
\begin{figure}[htb]
\begin{center}
\includegraphics[width=6.0cm,clip]{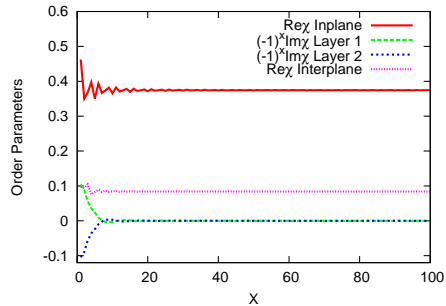}
\caption{(Color online) Spatial variations of the bond OPs for $\delta=0.15$.}
\end{center}
\end{figure}
%
%
%

The results for other values of $\delta$ show qualitatively the same
behavior; the absolute values of Im $\chi$ are larger (smaller) for 
smaller (larger) $\delta$, and the surface flux state persists to 
$\delta \sim 0.3$. 
In MF calculations for uniform systems, the type B flux phase arises for 
$\delta \lesssim 0.15$, and the transition to the type A state occurs 
as $\delta$ increases.  
The latter state persists to $\delta = \delta_c \sim 0.2$ 
in uniform systems.\cite{KK3}   
On the contrary, in the present BdG calculation, only the type B phase occurs,  
and $\delta_c $ is much larger ($\delta_c \sim 0.3$) than that in 
the uniform case. 
This is because the incommensurate flux phase, which is not taken 
into account in the uniform case, may be possible in nonuniform cases, 
and the type B incommensurate flux state has free energy lower 
than that of the type A  incommensurate state. 

For larger $J_\perp$, $\chi^\perp_j$ may be a complex number.\cite{Zhao}
However, $J_\perp$ in that case should be unrealistically large, 
and $\chi^\perp_j$ is real for the parameters appropriate for YBCO. 

In BdG calculations, the type A flux phase may be obtained as a metastable state 
that has free energy higher than that of the type B state. 
In this state, Im$\Delta_s$ in two layers have the same sign, in contrast 
to the case of type B phase.
This indicates that Im$\Delta_s$ is induced by Im $\chi$, and 
its sign is determined by the latter. 
In the type A state, the imaginary part of the interlayer pairing OP, 
Im$\Delta^\perp$, is finite. 
Since $\Delta^\perp$ has the same symmetry as the inplane $s$-wave SCOP, 
$\Delta_s^{(i)}$,  there is a bilinear coupling term in Ginzburg-Landau free energy, 
$\gamma (\Delta_s^{(1)} + \Delta_s^{(2)}) \Delta^\perp$, with $\gamma$  
being a coupling coefficient. 
This induces Im$\Delta^\perp$ once Im$\Delta_s^{(i)}$ becomes finite. 
In the type B phase, however, $\Delta_\perp$ vanishes 
because $\Delta_s^{(1)} = -\Delta_s^{(2)}$.

Next we study the LDOS. 
The LDOS at the $j$ site of the layer 1 is given as 
\begin{equation}
\begin{array}{rl}
N_1(j,E) =   & \displaystyle  \frac{1}{N_y} \sum_{k,n}
\Big(|u_{4j-3,n}(k)|^2 \delta(E-E_n(k)) \\
& + |u_{4j-2,n}(k)|^2 \delta(E+E_n(k))\Big),  
\end{array}
\end{equation}
and the LDOS for the layer 2 is obtained by replacing the subscripts, 
$(4j-3) \to (4j-1)$, $(4j-2) \to (4j)$. 
In numerical calculations we replace the $\delta$ function by 
a Lorentzian with the width $0.01J$. 
In Figs. 3-5, the LDOS at the surface and in the bulk are shown for 
$\delta=0.10$, $0.15$, and $0.20$.  
(The LDOS for the layer 1 and 2 are the same.) 
It is found that the splitting of peaks occurs in agreement 
with the experiment.\cite{Coving}
The height of the peaks become larger when $\delta$ gets smaller, while 
the peak splitting, $\Delta E$, changes only slightly in a nonmonotonic way;  
$\Delta E = 0.0763J, 0.0903J$, and $0.0777J$ 
for $\delta = 0.10, 0.15$, and $0.20$, respectively.  

%
\begin{figure}[htb]
\begin{center}
\includegraphics[width=6.0cm,clip]{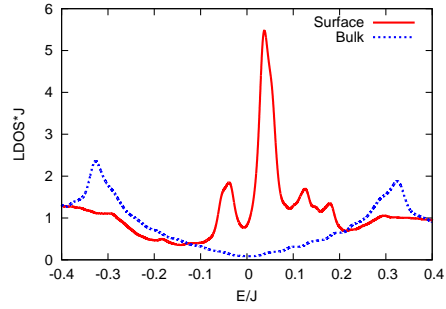}
\caption{(Color online) LDOS at the surface and in the bulk for $\delta=0.10$.}
\end{center}
\end{figure}
%
\begin{figure}[htb]
\begin{center}
\includegraphics[width=6.0cm,clip]{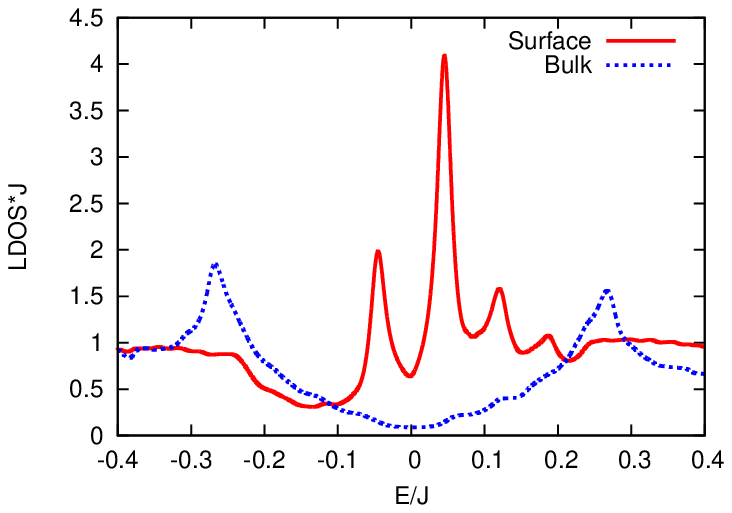}
\caption{(Color online) LDOS at the surface and in the bulk for $\delta=0.15$.}
\end{center}
\end{figure}
%
\begin{figure}[htb]
\begin{center}
\includegraphics[width=6.0cm,clip]{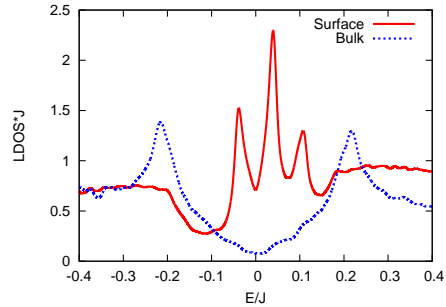}
\caption{(Color online) LDOS at the surface and in the bulk for $\delta=0.20$.}
\end{center}
\end{figure}
%

In order to understand the physical origin of the peak splitting in this model, 
we show the LDOS at the surface of a state with only $d$-wave SC order, 
and that with a surface flux phase as well as bulk $d$-wave SC order 
({\it i.e.}, without $s$-wave SCOP) in Fig.6.
Here the parameters are the same as those used in Fig.3. 
It is seen that the peak splitting occurs as long as the flux phase is present. 
This indicates that the flux-phase order, not the second SCOP Im$\Delta_s$, 
is the necessary ingredient for the peak splitting. 
We note that it is not possible to have a state with an $s$-wave SCOP
without the flux phase in the present model. 
Next we show the LDOS of a state with bulk (metastable) flux-phase order 
(type B) in Fig.7. All SCOPs are set to be zero, 
and the parameters are the same as in Fig.3. 
The LDOS in the bulk has broad peaks, 
and one of the peaks shifts near to $E=0$ at the surface. 
By comparing Figs. 6 and 7 with Fig.3,  we can see that 
the peak structure of the latter is mainly due to the flux phase, 
and the $d$-wave SC order also contributes to the behavior of the LDOS.

\begin{figure}[htb]
\begin{center}
\includegraphics[width=6.0cm,clip]{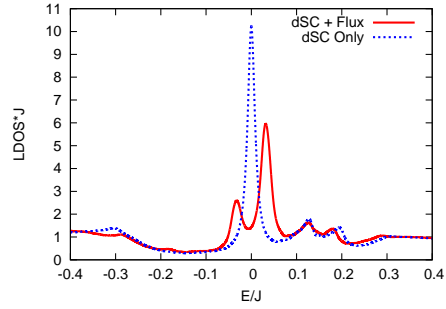}
\caption{(Color online) LDOS at the surface of a state without $s$-wave SCOP 
for $\delta=0.10$.}
\end{center}
\end{figure}

\begin{figure}[htb]
\begin{center}
\includegraphics[width=6.0cm,clip]{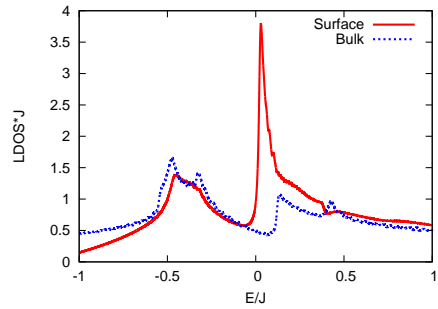}
\caption{(Color online) LDOS at the surface and in the bulk for a flux phase 
with $\delta=0.10$. Here all SCOPs are set to be zero}
\end{center}
\end{figure}

\section{Summary}
We have studied the states near the (110) surface of high-$T_C$ cuprate 
YBCO that are described by the bilayer $t-J$ model. 
Near the surface, superconductivity is strongly suppressed, and the flux phase 
in which the directions of the flux in two layers are opposite may occur 
in a wide doping region. 
Then ${\cal T}$ symmetry is violated and the LDOS at the surface has split peaks 
consistent with experimental findings.\cite{Coving} 
The spontaneous magnetic field that could arise near the surface with 
${\cal T}$ violation will be very small,  because the contributions from two layers 
essentially cancel out each other. 
These results may explain why no magnetic field is observed in some 
experiments for (110) surfaces of YBCO, for which the sign of 
${\cal T}$ violation is detected. 


\begin{acknowledgment}
 The author thanks M. Hayashi and H. Yamase for useful discussions. 
This work was supported by JSPS KAKENHI Grant Number 24540392.
\end{acknowledgment}



\begin{thebibliography}{9}

\bibitem{Coving} M. Covington, M. Aprili, E. Paraoanu, L. H. Greene, F. Xu, J.
Zhu, and C. A. Mirkin, Phys. Rev. Lett. {\bf 79}, 277 (1997). 

\bibitem{TRSB1}  
J. Xia, E. Schemm, G. Deutscher, S. A. Kivelson, D. A. Bonn, 
W. N. Hardy, R. Liang, W. Siemons, G. Koster, M. M. Fejer, and A. Kapitulnik, 
Phys. Rev. Lett. {\bf 100}, 127002 (2008).

\bibitem{TRSB2} H. Karapetyan, M. H\"ucker, G. D. Gu, J. M. Tranquada, M. M. Fejer, 
J. Xia, and A. Kapitulnik, Phys. Rev. Lett. {\bf 109}, 147001 (2012). 

\bibitem{TRSB3} H. Karapetyan, J. Xia, M. Hucker, G. D. Gu, J. M. Tranquada, 
M.M. Fejer, A. Kapitulnik, Phys. Rev. Lett. {\bf 112}, 047003 (2014). 

\bibitem{Fogel} M. Fogelstr\"om, D. Rainer, and J. A. Sauls, 
Phys. Rev. Lett. {\bf 79}, 281 (1997). 

\bibitem{Matsu1} M. Matsumoto and H. Shiba, J. Phys. Soc. Jpn. 
{\bf 64}, 3384 (1995).

\bibitem{Matsu2} M. Matsumoto and H. Shiba, J. Phys. Soc. Jpn. 
{\bf 64}, 4867 (1995). 

\bibitem{Carmi} R. Carmi, E. Polturak, G. Koren, and A. Auerbach, 
Nature {\bf 404}, 853 (2000). 

\bibitem{msr} H. Saadaoui, Z. Salman, T. Prokscha, A. Suter, H. Huhtinen, 
P. Paturi, and E. Morenzoni, Phys. Rev. B{\bf 88}, 180501(R) (2013). 

\bibitem{KK1} K. Kuboki, J. Phys. Soc. Jpn. {\bf 83}, 015003 (2014). 

\bibitem{KK2}  K. Kuboki, J. Phys. Soc. Jpn. {\bf 83}, 054703 (2014). 

\bibitem{Affleck} I. Affleck and J. B. Marston, 
Phys. Rev. B{\bf 37}, 3774 (1988). 

\bibitem{Zhang} F. C. Zhang, Phys. Rev. Lett. {\bf 64}, 974 (1990). 

\bibitem{Hamada} K. Hamada and D. Yoshioka, Phys. Rev. B{\bf 67}, 
 184503 (2003).

\bibitem{Bejas} M. Bejas, A. Greco, and H. Yamase,  Phys. Rev. B{\bf 86}, 
 224509 (2012).
 
 \bibitem{Zhao} H. Zhao and J. R. Engelbrecht, Phys. Rev. B {\bf 71}, 
054508 (2005). 
 
 \bibitem{KK3} K. Kuboki, J. Phys. Soc. Jpn. {\bf 83}, 125001 (2014).  

\bibitem{dG} P. G. de Gennes, {\it Superconductivity of Metals and Alloys}
(Addison-Wesley, Reading, MA, 1989).

\bibitem{Ogata} For a review on the $t-J$ model, see 
M. Ogata and H. Fukuyama, 
Rep. Prog. Phys. {\bf 71}, 036501 (2008). 

\bibitem{Andersen}  O. K. Andersen, A. I. Lichtenstein, O. Jepsen, and F. Paulsen, 
J. Phys. Chem. Solids {\bf 56}, 1573 (1995). 

 \bibitem{Zou} Z. Zou and P. W. Anderson, Phys. Rev. B{\bf 37}, 627 (1988). 

\bibitem{Lee} P. A. Lee, N. Nagaosa, and X.-G. Wen,  
Rev. Mod. Phys. {\bf 78}, 17 (2006). 

\bibitem{Kotliar} G. Kotliar and J. Liu, Phys. Rev. B {\bf 38}, 5142 (1988).

\bibitem{Suzumura} Y. Suzumura, Y. Hasegawa, and H. Fukuyama, 
J. Phys. Soc. Jpn. {\bf 57}, 2768 (1988).

\bibitem{Zhang2} F. C. Zhang, C. Gros, T. M. Rice, and H. Shiba, 
Supercond. Sci. Technol. {\bf 1}, 36 (1988). 

\bibitem{KKBdG} K. Kuboki, J. Phys. Soc. Jpn. {\bf 68}, 3150 (1999).

\bibitem{Tanuma} Y. Tanuma, Y. Tanaka, M. Ogata, and S. Kashiwaya, 
Phys. Rev. B {\bf 60}, 9817 (1999).

\bibitem{Zhu} J. X. Zhu and C. S. Ting, 
Phys. Rev. B {\bf 61}, 1456 (2000).

\bibitem{KKBdG2} K. Kuboki and H. Takahashi, 
Phys. Rev. B {\bf 70}, 214524 (2004). 

\bibitem{Yamase} H. Yamase and W. Metzner, Phys. Rev. B{\bf 73}, 214517 (2006). 

\end{thebibliography}
\end{document}